\documentclass[reprint, superscriptaddress, amsmath,amssymb, aps, pra, floatfix]{revtex4-1}
\usepackage[T1]{fontenc} 
\usepackage{mathptmx}
\usepackage{graphicx}
\usepackage{dcolumn}
\usepackage{bm}
\usepackage[colorlinks, breaklinks, citecolor=blue,linkcolor=black,urlcolor=black]{hyperref}

\begin{document}

\title{Dual-species Optical tweezer for Rb and K atoms}

\author{Yangbo Wei}
\author{Kedi Wei}
\author{Shangjin Li}
\affiliation{%
Zhejiang Key Laboratory of Micro-nano Quantum Chips and Quantum Control, School of Physics, and State Key Laboratory for Extreme Photonics and Instrumentation, Zhejiang University, Hangzhou 310027, China
}%
\author{Bo Yan}
\email{yanbohang@zju.edu.cn}
\affiliation{%
Zhejiang Key Laboratory of Micro-nano Quantum Chips and Quantum Control, School of Physics, and State Key Laboratory for Extreme Photonics and Instrumentation, Zhejiang University, Hangzhou 310027, China
}%
\affiliation{%
College of Optical Science and Engineering, Zhejiang University, Hangzhou 310027, China
}%

\date{\today}

\begin{abstract}
The optical tweezer experiment with neutral atoms is a focal topic in cold atom physics due to its significant potential in quantum computing and simulation. Here, we present the realization of a dual-species optical tweezer for both Rb and K atoms, marking the first step towards creating a polar molecule optical tweezer array. Initially, Rb and K atoms are collected using a dual magneto-optical trap (MOT) and further cooled to 7 $\mu$K for Rb and 10 $\mu$K for K. By employing 850 nm tweezer beams, we demonstrate the ability to capture individual Rb or K atoms. The filling ratios of Rb and K can be finely adjusted by controlling the atomic densities of both species. Utilizing the post-selection technique, we can create a deterministic array of two-species atoms, paving the way for future polar molecule array formation.
\end{abstract}

\maketitle

\section{introduction}
 
	The optical tweezers with neutral atoms provide powerful platforms for quantum information processing \cite{Schlosser2001, Zhang2010, Wilk2010, Kaufman2012, Lester2015, Kaufman2021}. These tweezers allow for the highly controllable sampling of atoms \cite{Barredo2018}. With advancements in post-selection techniques \cite{Weiss2004, Barredo2016, Endres2016, Sheng2021}, the fidelity of capturing a single atom in each tweezer approaches unity, significantly increasing the loading rate fidelity in the tweezer array. To date, several hundred to even thousands of atoms have been successfully trapped in optical tweezer arrays \cite{Tian2023, Tao2023, Huft2022, Manetsch2024, Lorenzo2021,manetsch2024tweezer}. Inspired by these achievements, researchers have simulated fascinating phenomena such as many-body physics \cite{Bernien2017, Ebadi2021}, and quantum magnetism using optical tweezer platforms \cite{Scholl2021}. Error correction computing has also been demonstrated with 48 logical qubits \cite{Bluvstein2024}, highlighting the potential of this technology in quantum computing. 
Recently, the application of optical tweezers has expanded into the field of laser cooling of cold molecules \cite{Anderegg2019}, and the spin exchange interaction of polar molecules has been observed in the optical tweezers\cite{Bao2023, Holland2023}.

Another fascinating development associated with optical tweezers is the creation of ultracold molecules one by one. The stimulated Raman adiabatic passage (STIRAP) technique enables the creation of ultracold polar molecules from ultracold atoms, such as KRb \cite{Ni2008, Tim2024}, RbCs \cite{Takekoshi2014}, NaK \cite{Park2015} and NaRb \cite{Guo2016, Ye_2018, Guo2018}. Using optical tweezers, researchers can trap pairs of these alkali atoms, merge them, and transfer them to the molecular ground state via STIRAP. These highly controlled single-molecule arrays are extremely useful for applications in quantum chemistry and quantum computing. Utilizing this strategy, researchers have successfully created individual ground-state polar molecules such as NaCs \cite{Zhang2020, Cairncross2021}, and RbCs \cite{Ruttley2023}.  Recently, precise control of the internal state of molecules has been achieved \cite{Cornish2024, Ni2024, Picard2023}.

 Among ultracold polar molecules, KRb was the first to be created in the ultracold regime \cite{Ni2008} and reached quantum degeneracy \cite{DeMarco2019}. However, the creation of single molecules remains a challenge. Here, we report the realization of a dual-species optical tweezer for Rb and K. Unlike setups for Na/Cs and Rb/Cs, we use a single laser to create the optical tweezer for both species, rather than employing a species-selective optical tweezer \cite{Liu2018, Brooks2021, Hannes2022}. By fine-tuning the density of the magneto-optical trap (MOT), we have achieved nearly equal loading rates for both Rb and K. By introducing a tweezer array and utilizing post-selection techniques, we can deterministically select Rb/K atom pairs with high fidelity. This capability is crucial for the further creation of a polar molecule array.

\section{experimental setup}\label{sec3}
\subsection{Dual MOT for Rb and K}
\begin{figure*}[]
\centering
\vspace{2mm}
\includegraphics[width=1\textwidth]{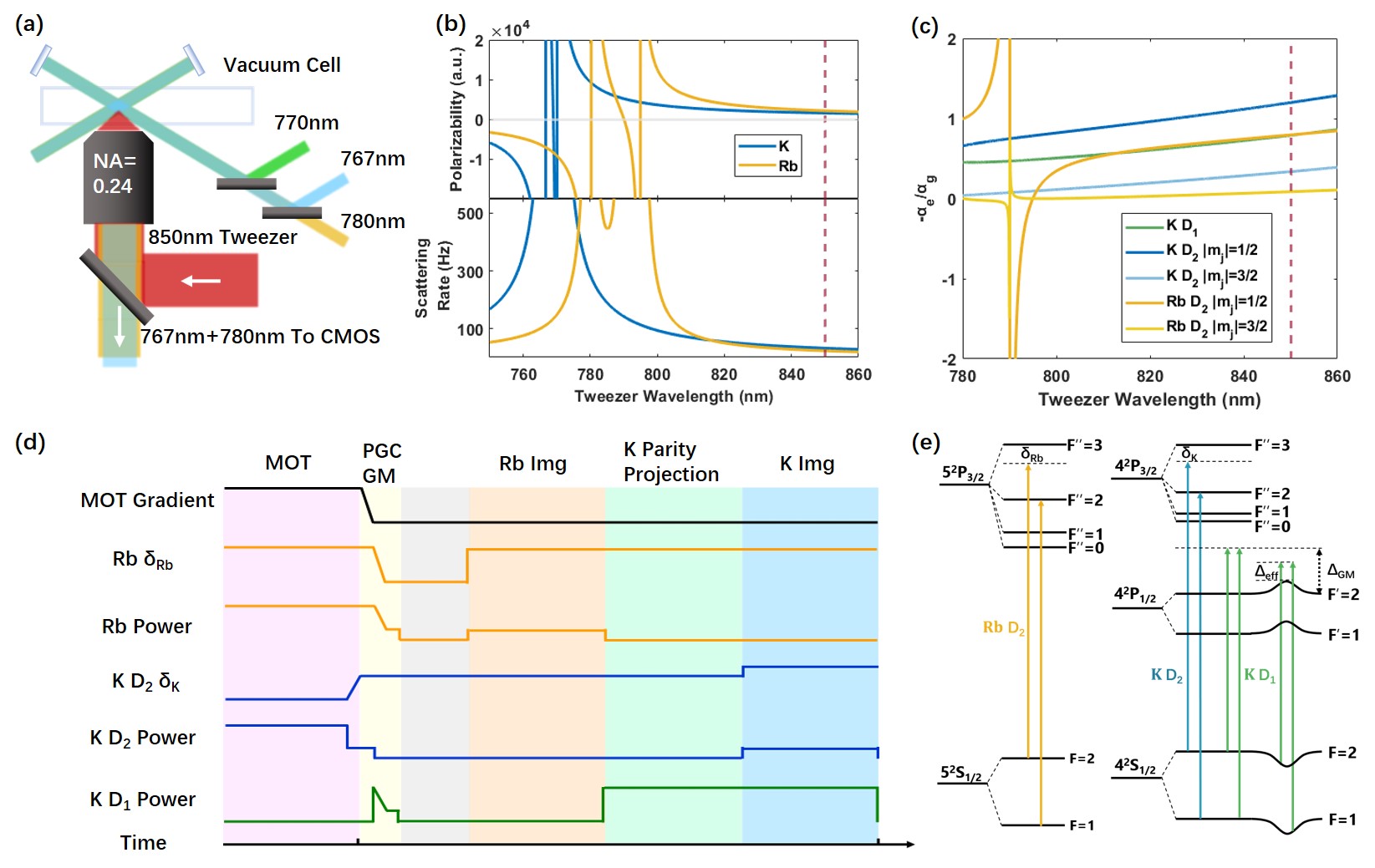}
\caption{ \label{setup}
(a) The experimental setup of the dual-species optical tweezer. A dual MOT of Rb and K is created for optical tweezer loading. The K and Rb MOT beams and the K $D_1$ gray molasses beams are combined with dichroic mirrors. The optical tweezers are formed with 850~$nm$ lasers. The objective lens has a $NA=0.24$. (b) Numerical results of the polarizability and scattering rate (with a trap depth of 1~mK). The red line indicates the wavelength (850~nm) used in our experiment, where Rb and K have similar polarizabilities, and the scattering rate is still low enough for optical tweezer loading. (c) Numerical results of the polarizability ratio between the excited state and the ground state. (d) The time sequence of our experiment. Rb and K are first collected with the dual MOT, then they are further cooled with PGC. In addition, we apply the $D_1$ GM for K atoms to further cool them down to tens $\mu$K. Finally, a dual-species detection is applied. (e) The relevant transition for Rb and K. The detunings are defined accordingly in the plot.
}
\end{figure*}

The experimental setup is depicted in Fig. \ref{setup} (a). Dual MOTs for both $^{87}$Rb and $^{39}$K are realized within a 10~mm$\times$ 20~mm$\times$ 80~mm vacuum glass cell. The background vacuum pressure reaches $1\times 10^{-9}~$Pa, monitored by the ion pump. For the Rb MOT, the cooling laser has a total power of 8~mW and is detuned by $-13$~MHz from the $D_2$ cycling transition, while the repump laser has a power of 1~mW. For the K MOT, the cooling laser power is 12~mW with a detuning of $-33$~MHz from the $D_2$ cycling transition, and the repump laser power is 8~mW. All these laser beams are overlapped and have a diameter of 5~mm.

 Figure \ref{setup} (d) shows the typical time sequence of our experiment. The MOT loading duration is set to be 500 ms, during which approximately $1\times 10^6$ atoms for both Rb and K are collected.  After that, {optical molasses}  is applied to both species. The magnetic gradient field is decreased from 11.5~G/cm to zero in 4~ms. The detuning of the Rb cooling laser, $\delta_{Rb}$ is ramped from $-13$~MHz to $-60$~MHz, and power is reduced from 8~mW to 200~$\mu$W in 13~ms. This {possesses a strong polarization gradient cooling (PGC) effect for Rb atoms and} cools them from 100$~\mu$K to 7(1)$~\mu$K. Simultaneously, the detuning of the K cooling laser, $\delta_{K}$ jumps from $-33$~MHz to $-7$~MHz, and the power is switched from 12~mW to 240~$\mu$W. {This process cools the K atoms from roughly 3$~mK$ to 48(5)$~\mu K$}, which is still too high for efficient loading into the optical tweezer. Fig. \ref{setup} (e) illustrates the laser detuning of Rb and K during the PGC process.

To further cool the K atoms,  a $\Lambda$-enhanced gray molasses (GM) relevant to the $D_1$ transition (770$~nm$) is applied, as indicated by the green lines in Fig. \ref{setup} (e) \cite{Ang'ong'a2022}. {This technique is particularly effective for cooling light alkali metals, and a recent publication \cite{Hood2023} demonstrates its effectiveness in the lightest alkali metal system, $^6Li$. }This $D_1$ gray molasses laser is modulated by a fiber electro-optic modulator (EOM) with 461.7 MHz, has a total power of 4~mW and a detuning of $\Delta_{GM}=+37$~MHz in free space. The beam path overlaps with the MOT beams, as shown in Fig. 1(a).  This process is highly effective and reduces the temperature of K atoms to 10(2)$~\mu$K within 10~ms. {Notably, the temperature measured in tweezer is around 30$\mu K$ for Rb and 45$\mu K$ for K, with the method of release and recapture \cite{Tuchendler2008}. The Lamb-Dicke parameters in our system, around 0.5 for Rb and 0.7 for K, are relatively high. This means that the motional state is not well reserved during the recoil from photon emission or absorption, resulting in an in-trap temperature higher than that in free space, which is consistent with a recent theoretic prediction\cite{Phatak2024}.}

\subsection{A Dual-species Optical Tweezer}

To create single polar molecule arrays, the first step is to build optical tweezers capable of trapping dual species.
In the previous dual-species tweezer experiment, two species atoms are trapped with species-selective loading \cite{Liu2018, Brooks2021, Hannes2022}. This method involves choosing the wavelength of the optical tweezer lasers between the $D$-lines of the two atoms, resulting in one species being trapped while the other is anti-trapped. In the case of Rb/Cs \cite{Brooks2021}, two lasers with wavelengths of $\lambda_{Rb}=814$~nm and $\lambda_{Cs}=938$~nm were selected. Here $\lambda_{Rb}$ lies between the $D_1$ wavelengths of Rb (795~nm) and Cs (894.6~nm), thereby forming a tweezer that only traps Rb. On the other hand, $\lambda_{Cs}$ is far from the $D$-lines of Rb, creating a trap that is shallow for Rb but sufficiently deep for Cs. This species-selective scheme works very well for Rb/Cs and Na/Cs pairs because the $D$-lines for those atom pairs are significantly different in wavelength. Choosing proper wavelengths makes it possible to achieve both high trap depth and low heating rates simultaneously.

However, the situation is different for the K/Rb case, as the $D$-lines of these two atoms are relatively close (795~nm, 780~nm, 770~nm, and 767~nm). The previous strategy doesn't work here. 
 Figure \ref{setup} (b) shows the numerical result of the polarizability and scattering rate for both Rb and K in a 1~mK Tweezer. When the wavelength of the tweezer laser is chosen between 767~nm and 795~nm, the scattering rate becomes too high. For a tweezer with a trap depth of 1~mK, at least one species's scattering rate will exceed 400~Hz, resulting in heating rates higher than 60~$\mu$K/s for Rb or 140~$\mu$K/s for K. Such high heating rates make tweezer loading difficult and shorten the lifetime considerably.  

To address this, we choose a wavelength of 850~nm for the optical tweezer laser, which is far away from the $D$-line transitions of both Rb and K. A similar system of dual-species tweezer for $^{87}Rb/^{85}Rb$ has been realized before \cite{Zhan2022}, in which both $^{87}Rb$ and $^{85}Rb$ can be captured by 808~nm tweezer. As indicated in Fig. \ref{setup} (b), the polarizabilities of Rb {(2138$~a.u.$) and K (1552$~a.u.$)} are similar (where $1~a.u.=1.64878\times 10^{-41}~J(V/m)^{-2}$), and the scattering rate is low enough ($<$30~Hz for both Rb and K in 1~mK tweezer).

{ On the other hand, the polarizability ratio between the excited and ground state $\alpha_e/\alpha_g$ is very important. A dipole trap for the ground state usually holds an anti-trapping effect for the excited state. During the image process, atoms spend a large amount of time in the excited state, and the anti-traping effect can reduce the overall trap force significantly and cause serious loss \cite{Hutzler_2017}. It is desirable to have a ratio of $-\alpha_e/\alpha_g$ around one. Fig. \ref{setup}(c) shows the ratio of $-\alpha_e/\alpha_g$ versus trap wavelength for both Rb and K. When the tweezer wavelength is 850 nm, the highest $-\alpha_e/\alpha_g$ for certain sublevels is around one, making it an appropriate choice.}

The 850~nm tweezer laser is focused by a homemade lens ($NA=0.24$), resulting in a waist of $w=2.1(1)~\mu$m at focus. The power in the cell is $P=$14~mW, which corresponds to a trap depth of 0.96~mK for Rb and 0.72~mK for K. The optical tweezer is turned on at the beginning of the MOT loading stage. %

\subsection{Dual-species detection}

\begin{figure}[tbp]
\centering
\includegraphics[width=0.48\textwidth]{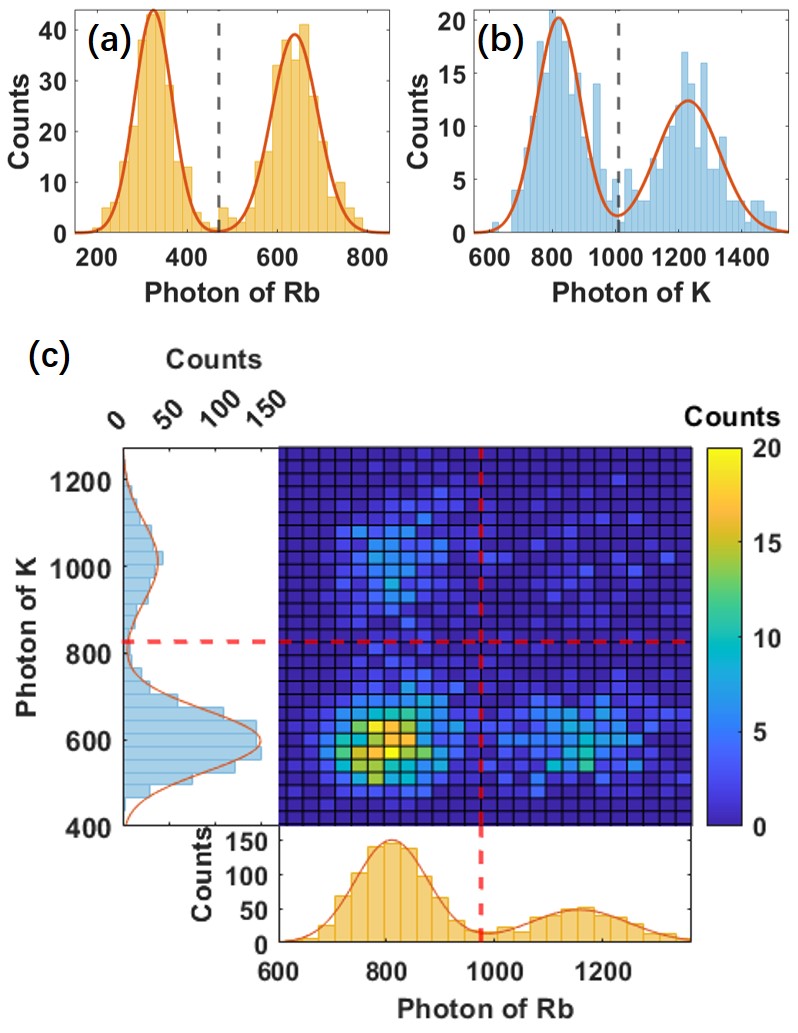}
\caption{ \label{detection}
The optical tweezer detection with (a) Rb and (b) K only. (c) The dual-species detection when both Rb and K are loaded into the optical tweezer. Two images with Rb and K are recorded successively. The $x$ and $y$ axes of the two-dimensional plot are Rb and K fluorescent counting numbers, respectively. The color bar indicates the statistic counting of Rb and K. {The dashed line represents the threshold we use to identify whether the tweezer is loaded or empty \cite{manetsch2024tweezer}, with a fidelity of $99\%$ for Rb and $98\%$ for K.}
}
\end{figure}

After loading Rb and K into the optical tweezer, we need to detect them. In the experiment, all the cooling lasers are turned off for 50~ms after PGC and GM, allowing atoms outside the tweezer to disperse and thus reducing the background during detection.

During the detection, we turn the cooling and repump lasers back on. For Rb, the total intensity of the cooling laser is about 19$I_\text{sat}$, here $I_\text{sat}$ is the saturation intensity, and the effective detuning is set $-37$~MHz (including the light shift of the optical tweezer).  For the repump laser, the effective detuning is $-24$~MHz. The illumination time is 100~ms. Figure \ref{detection} (a) shows the typical data of our single-Rb imaging, with a filling ratio of about 50$\%$.

For K detection, we turn on both molasses and GM lasers. If only the molasses laser were on, the cooling efficiency would be quite low, and the lifetime of K in the optical tweezer would be too short for effective detection. The GM laser provides efficient cooling and allows sufficient photon scattering before K atoms get lost. The 767 nm cooling lasers have a total intensity of 1$I_\text{sat}$ and are near-resonant with $|F=2\rangle$ to $|F''=3, m_F=\pm2\rangle$. Here the 767 nm K repump laser is not used since the GM laser can repump K atoms back to $|F=2\rangle$. Before imaging K, we turn on the GM laser for 100~ms to apply the parity projection, otherwise, more than one K atoms can survive in an optical tweezer. The typical filling ratio of K is about 47$\%$, as shown in Fig. \ref{detection} (b).

We apply dual-species detection to detect both Rb and K in the optical tweezer. After loading Rb/K atoms into the optical tweezer, we first initiate the Rb image process for 100~ms. Then, we apply the K parity projection for 100~ms. Finally, we perform the K image. Figure \ref{detection} (c) presents a two-dimensional plot of these results. The color bar indicates the statistic counting of Rb and K. The left and bottom sides of the plot show the statistical results for counting only K or Rb, respectively. In this scenario, both atoms exhibit a filling ratio of around 25$\%$. The two-dimensional plot delineates four areas. The bottom-left corner indicates no atoms are loaded in the tweezer. The top-right area shows both Rb and K are loaded into the trap, with an experimental filling ratio of about $4\%$ in this area. This number is consistent with no Rb/K can be loaded into the trap at the same time. The minor residual number might be attributed to our detection fidelity imperfections or insufficiently violent light-assisted collisions due to the relatively large trap waist. Additionally, we observe that sometimes atoms can be captured when the cooling laser is turned on, even without the MOT stage.

%

\begin{figure}[]
\centering
\includegraphics[width=0.48\textwidth]{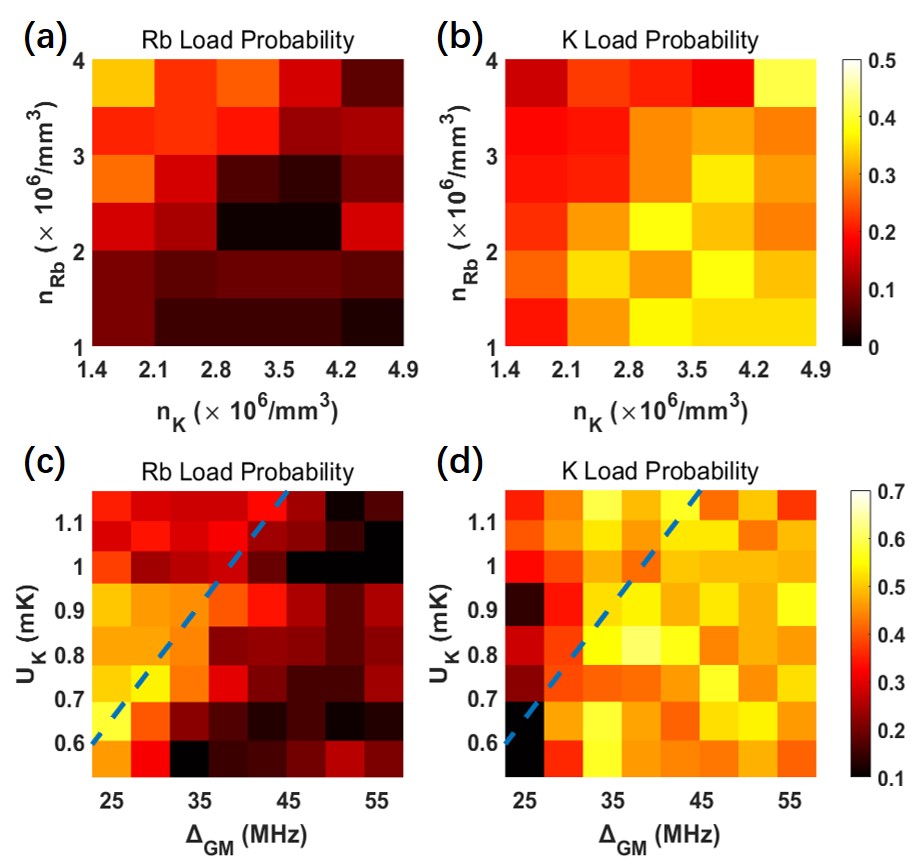}
\caption{ \label{loading}
 The optical tweezer loading rates of Rb (a) and K (b) are dependent on Rb MOT density $n_{Rb}$ and K MOT density $n_{K}$.  The Rb (c) and K (d) loading rate in the optical tweezer versus $\Delta_{GM}$ and $U_K$.
}
\end{figure}

To achieve a similar loading rate for both Rb and K, it is crucial to fine-tune the atomic density. In our experiment, we adjust the MOT density by tuning the MOT loading time. Figures \ref{loading} (a) and (b) show the results. Optimal Rb loading occurs when the Rb MOT density ($n_{Rb}$) is high and the K MOT density ($n_K$) is low. Conversely, optimal K loading occurs at high K MOT density and low Rb MOT density. These results indicate a strong anti-correlation in the filling ratio of Rb and K within a dual MOT, attributed to the photo--ssociated inelastic collisions between Rb and K \cite{Schlosser2002, Kuppens2000}.

Previous experiments have shown that optical tweezer loading is highly sensitive to the single-photon detuning $\Delta_{GM}$ of the gray molasses and the trap depth of K ($U_K$) the tweezer \cite{Brown2019, Ang'ong'a2022, Loh2021, Lester2015}. Figures \ref{loading}(c) and \ref{loading}(d) illustrate the tweezer filling ratios as functions of $U_K$ and $\Delta_{GM}$. The dashed line represents the condition where the induced AC-Stark light shift $\Delta_{AC}=\Delta_{GM}$. Above the dashed line, the effective detuning of GM ($\Delta_\text{eff} = \Delta_{GM} - \Delta_{AC}$, as shown in Fig. \ref{setup} (e)) is red, leading to atom heating and a reduced K loading rate. Our experiment does not show a sharp transition in the K loading rate across this line as observed in other studies \cite{Brown2019, Ang'ong'a2022}. We attribute this to the relatively large beam waist of our tweezer beam (2.1~$\mu$m). Although $\Delta_\text{eff}$ is red-detuned at the center of the optical tweezer, it remains blue-detuned at the edges, assisting in atom cooling. This effect is exacerbated with larger tweezer waists. Nevertheless, we still observe a significant change in the Rb loading rate across the dashed line. The sensitivity of K loading to $\Delta_{GM}$ is transferred to Rb due to the anti-correlation in the loading ratios of the two species.


\begin{figure}[t]
\centering
\includegraphics[width=0.48\textwidth]{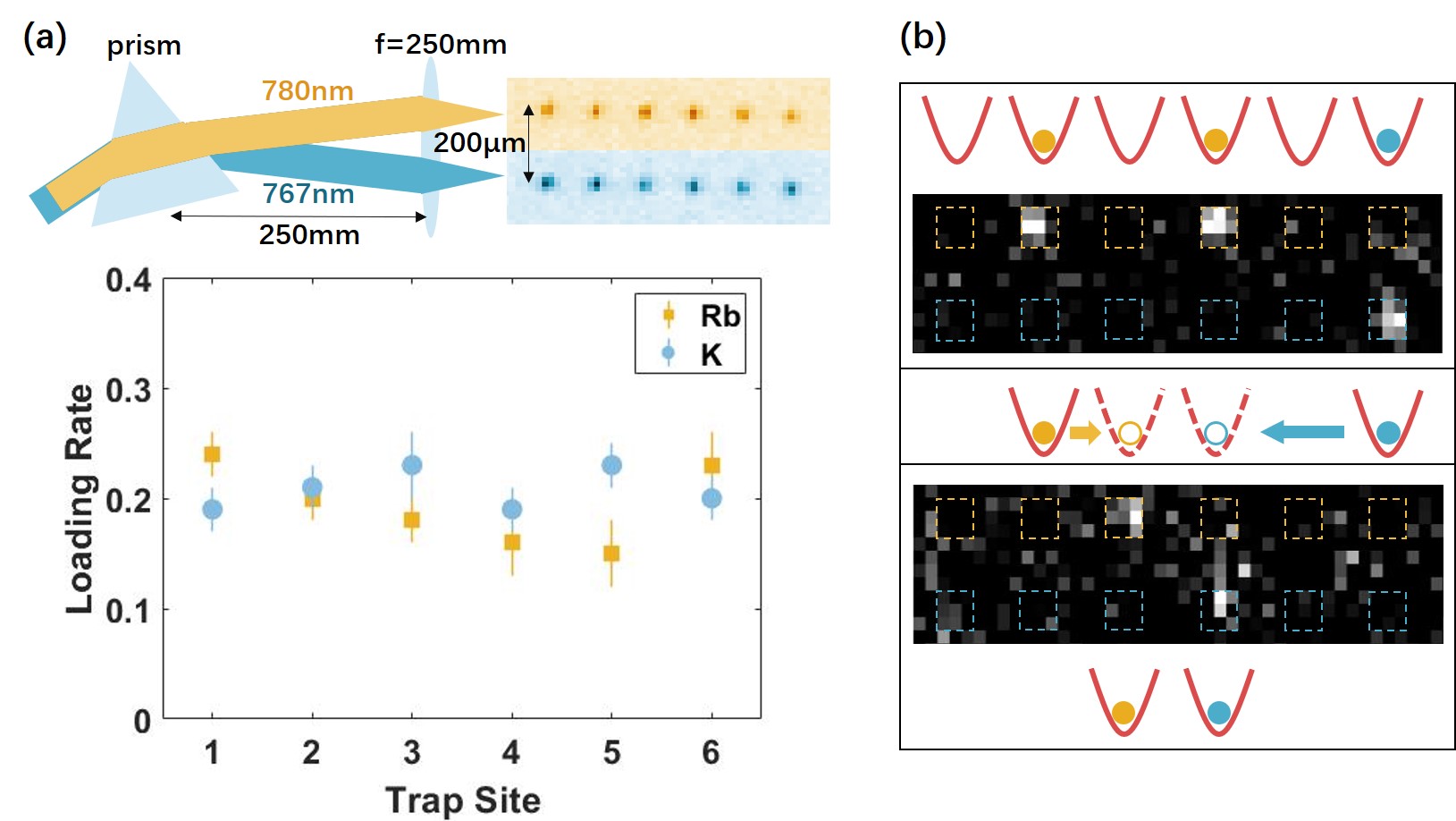}
\caption{ \label{array}
{ (a) Top: The prism-assistant image system for both Rb (yellow) and K (blue). The Rb and K images are separated by approximately $200\mu m$. 
Bottom: loading rate for Rb or K of each site of the dual-species tweezer array. (b) Rearrangement steps. Top: the initial loading state. Middle: one Rb and one K are moved to the target traps (here are trap-3 and trap-4). Bottom: the image after rearrangement. The up line is for Rb (yellow) and the down line is for K (blue).}
}
\end{figure}

\subsection{1D tweezer array rearrangement}\label{sec4}

The above experiment successfully loaded Rb and K into a dual-species optical tweezer. To form ultracold polar molecules, it is essential to load Rb/K atom pairs deterministically. An optical tweezer array combined with a post-selection technique can address this requirement \cite{Bernien2017, Browaeys2016}.

{While the two-time image of the dual-species detection works, it takes a relatively longer time. To shorten the cycle time and reduce the systematic detection error, we developed a new imaging system that can image both Rb and K simultaneously in different camera regions with the help of a prism, as shown in Fig. \ref{array} (a).
 After the prism, the 780~nm and 767 nm fluorescence are separated by an angle of $2.7'$ and then focused by a lens ($f=250$~mm), resulting in a separation of Rb and K images by approximately $200\mu m$ on the camera. The additional prism causes a loss of signal approximately $8\%$ for both species. In this way, we can obtain Rb/K loading state with one shot. With this detection method, we don't observe Rb and K signals in the same trap. Figure \ref{array} (a) top shows the average image of 200 shots with both Rb and K. Both Rb and K can be loaded into the tweezer array and the bottom figure shows the loading rate for both Rb and K in dual-species tweezer.}

Figure \ref{array} (b) illustrates the rearrangement process of the dual-species tweezer array. In our experiment, an acoustic-optic deflector (AOD) is used to generate a one-dimensional array with six tweezers. Each tweezer can capture either Rb or K, with a loading rate of $19\%$ for Rb and $21\%$ for K. Initially, atoms are randomly loaded into a tweezer array. After capturing the initial image, we identify which traps are loaded with Rb or K and shut down the empty traps. Then we transfer one Rb and one K to the target traps (trap-3 and trap-4 in the example). During the moving process, all the other traps on the road are turned off to avoid heating atoms, and the loss induced by moving is less than $2\%$ for both Rb and K. The entire moving process is completed within 20 ms. Finally, we image the atoms again to confirm successful relocation.  To avoid heating atoms in the merging process, an independent tweezer overlapped with trap-3 is set up as a transit tweezer. By employing post-selection and rearrangement techniques, the probability of finding one Rb/K atom pair in one sequence increases to $53\%$, providing a promising starting point for ultracold polar molecule tweezer experiments.

\section{conclusion}\label{sec5}
To conclude, we have successfully realized a dual-species optical tweezer setup capable of trapping both Rb and K atoms, overcoming the defect caused by their too-close $D$ line wavelength. Additionally, we extended the setup from a single tweezer to a one-dimensional tweezer array. Utilizing a post-selection method, we can deterministically load Rb/K pairs in the target tweezers, paving the way for further experiments with polar molecules. Notably, this method does not rely on species-selective wavelengths, making it applicable to various atom pairs and potentially even to three or more species. Right now, we still work with $^{39}$K because of the rich natural abundance. In the future, we plan to switch to $^{40}$K. 

Furthermore, we have developed a new detection method that allows us to capture images of Rb and K simultaneously using a prism to separate the fluorescent images on the camera. This adjustment saves us 100 ms, effectively shortening the cycle time. This method has promising applications in situations that require extremely short cycle times and is particularly suitable for dual-species experiments where the fluorescent wavelengths are different.

\begin{acknowledgments}

We thank Dr. Weikun Tian for the fruitful discussion. We acknowledge the support from the National Natural Science Foundation of China under Grant No. U21A20437 and No. 12074337, the National Key Research and Development Program of China under Grant No.2023YFA1406703 and No. 2022YFA1404203, and the Fundamental Research Funds for the Central Universities under Grant No. 226-2023-00131.
\end{acknowledgments}

\bibliographystyle{apsrev4-1}
\bibliography{tweezerbib}

\end{document}